\documentclass[12pt,tightenlines,eqsecnum,floats,aps,amsmath,amssymb,nofootinbib,prd]{revtex4}

\usepackage{amsmath,amssymb,amsfonts,latexsym}
\usepackage{graphicx}
\usepackage{enumerate} 
\usepackage{colordvi} 
\usepackage[mathscr]{eucal}
\usepackage{epstopdf}

\def\P{\mathbb{P}}

\def\R{\mathbb{R}}

\def\g{\gamma}

\def\a{\mathfrak{a}}
\def\W{\mathfrak{W}}
\def\C{\mathfrak{C}}

\def\be{\begin{equation}}
\def\ee{\end{equation}}
\def\ba{\begin{eqnarray}}
\def\ea{\end{eqnarray}}

\def\A{{\cal A}}
\def\Ab{\bar{\cal A}}
\def\C{{\cal C}}

\def\H{{\cal H}}

\def\Hp{\H_{\rm phy}}

\def\P{{\cal P}}

\def\Ab{{\bar \A}}
\def\h{\hat }

\def\lp{\ell_{\rm Pl}}
\def\M{M}
\def\SU(2){{\rm SU(2)}}
\def\su(2){{\rm su(2)}}

\def\Cyl{{\rm Cyl}}
\def\cyl{\Cyl}
\def\su{{\rm su}}
\def\SU{{\rm SU}}

\def\Tr{{\rm Tr\,}}

\def\lp{{\ell}_{\rm Pl}}

\def\e{{}^o\!e}
\def\w{{}^o\!\omega}
\def\grav{{\rm grav}}

\newcommand{\ket}[1]{\ensuremath{|#1\rangle}}
\newcommand{\ip}[2]{{\langle#1\,|\,#2\rangle}}

\usepackage{enumerate}

\usepackage{colordvi}

\def\f{\frac}

\def\dd{\textrm{d}}
\def\ub{\underbar}
\def\ul{\underline}

\def\t{\tilde}
\def\p{\partial}

\usepackage[mathscr]{eucal}

\def\lp{\ell_{\rm Pl}}
\def\ps{\mathbf{\Gamma}}
\def\P{\mathbb{P}}

\def\R{\mathbb{R}}
\def\M{M}
\def\Tr{{\rm Tr}}
\def\SU(2){{\rm SU(2)}}
\def\su(2){{\rm su(2)}}
\def\U(1){{\rm U(1)}}
\def\A{{\cal A}}
\def\Ab{\bar{\cal A}}
\def\H{{\cal H}}

\begin{document}

\title{Gravity, Geometry and the Quantum}

\author{Abhay Ashtekar${}^{1,2}$}
\affiliation{Institute for Gravitational Physics and Geometry \\
Physics Department, Penn State, University Park, PA 16802-6300 \\
${}^2$ Institute for Theoretical Physics, University of Utrecht,
Princetonplein5, 3584 CC Utrecht, The Netherlands}

\begin{abstract}

After a brief introduction, basic ideas of the quantum Riemannian
geometry underlying loop quantum gravity are summarized. To
illustrate physical ramifications of quantum geometry, the
framework is then applied to homogeneous isotropic cosmology.
Quantum geometry effects are shown to replace the big bang by a
big bounce. Thus, quantum physics does not stop at the big-bang
singularity. Rather there is a pre-big-bang branch joined to the
current post-big-bang branch by a `quantum bridge'. Furthermore,
thanks to the background independence of loop quantum gravity,
evolution is deterministic across the bridge.

\end{abstract}
\maketitle

\section{INTRODUCTION}
\label{s1}

General relativity and quantum theory are among the greatest
intellectual achievements of the 20th century. Each of them has
profoundly altered the conceptual fabric that underlies our
understanding of the physical world. Furthermore, each has been
successful in describing the physical phenomena in its own domain
to an astonishing degree of accuracy. And yet, they offer us {\it
strikingly} different pictures of physical reality. Our past
experience in physics tells us that these two pictures must be
approximations, special cases that arise as appropriate limits of
a single, universal theory.  That theory must therefore represent
a synthesis of general relativity and quantum mechanics. This
would be the quantum theory of gravity that we invoke when faced
with phenomena, such as the big bang and the final state of black
holes, where the worlds of general relativity and quantum
mechanics must unavoidably meet.

Remarkably, the necessity of a quantum theory of gravity was
pointed out by Einstein already in 1916. In a paper in the
Preussische Akademie Sitzungsberichte he wrote:
\begin{quote}
\textsl{Nevertheless, due to the inneratomic movement of
electrons, atoms would have to radiate not only electromagnetic
but also gravitational energy, if only in tiny amounts. As this is
hardly true in Nature, it appears that quantum theory would have
to modify not only Maxwellian electrodynamics but also the new
theory of gravitation.}
\end{quote}
Ninety years later, our understanding of the physical world is
vastly richer but a fully satisfactory unification of general
relativity with quantum physics still eludes us. Indeed, the
problem has now moved to the center-stage of fundamental physics.
(For a brief historical account of the evolution of ideas see,
e.g., \cite{aanjp}.)

A key reason why the issue is still open is the lack of
experimental data with direct bearing on quantum gravity. As a
result, research is necessarily driven by theoretical insights on
what the key issues are and what will `take care of itself' once
this core is understood.  As a consequence, there are distinct
starting points which seem natural. Such diversity is not unique
to this problem. However, for other fundamental forces we have had
clear-cut experiments to weed-out ideas which, in spite of their
theoretical appeal, fail to be realized in Nature. We do not have
this luxury in quantum gravity. But then, in absence of strong
experimental constraints, one would expect a rich variety of
internally consistent theories. Why is it then that we do not have
a single one? The reason, I believe, lies the deep conceptual
difference between the description of gravity in general
relativity and that of non-gravitational forces in other
fundamental theories. In those theories, space-time is given a
priori, serving as an inert background, a stage on which the drama
of evolution unfolds. General relativity, on the other hand, is
not only a theory of gravity, it is also a theory of space-time
structure. Indeed, in general relativity, gravity is encoded in
the very geometry of space-time. Therefore, a quantum theory of
gravity has to simultaneously bring together \emph{gravity,
geometry and the quantum}. This is a band new adventure and our
past experience with other forces can not serve as a reliable
guide.

Loop quantum gravity (LQG) is an approach that attempts to face
this challenge squarely (for details, see, e.g.,
\cite{alrev,crbook,ttbook}). Recall that Riemannian geometry
provides the appropriate mathematical language to formulate the
physical, kinematical notions as well as the final dynamical
equations of any classical theory of relativistic gravity. This
role is now assumed by \textit{quantum} Riemannian geometry. Thus,
in LQG both matter and geometry are quantum mechanical `from
birth'.

In the classical domain, general relativity stands out as the best
available theory of gravity. Therefore, it is natural to ask:
\textit{Does quantum general relativity, coupled to suitable
matter} (or supergravity, its supersymmetric generalization)
\textit{exist as consistent theories non-perturbatively?} In
particle physics circles the answer is often assumed to be in the
negative, not because there is concrete evidence which rules out
this possibility, but because of the analogy to the theory of weak
interactions. There, one first had a 4-point interaction model due
to Fermi which works quite well at low energies but which fails to
be renormalizable. Progress occurred not by looking for
non-perturbative formulations of the Fermi model but by replacing
the model by the Glashow-Salam-Weinberg renormalizable theory of
electro-weak interactions, in which the 4-point interaction is
replaced by $W^\pm$ and $Z$ propagators. It is often assumed that
perturbative non-renormalizability of quantum general relativity
points in a similar direction. However this argument overlooks a
crucial and qualitatively new element of general relativity.
Perturbative treatments pre-suppose that space-time is a smooth
continuum \textit{at all scales} of interest to physics under
consideration. This assumption is safe for weak interactions. In
the gravitational case, on the other hand, the scale of interest
is \emph{the Planck length} and there is no physical basis to
pre-suppose that the continuum approximation should be valid down
to that scale. The failure of the standard perturbative treatments
may largely be due to this grossly incorrect assumption and a
non-perturbative treatment which correctly incorporates the
physical micro-structure of geometry may well be free of these
inconsistencies.

Are there any situations, outside LQG, where such physical
expectations are borne out by detailed mathematics? The answer is
in the affirmative. There exist quantum field theories (such as
the Gross-Neveu model in three dimensions) in which the standard
perturbation expansion is not renormalizable although the theory
is \emph{exactly soluble}! Failure of the standard perturbation
expansion can occur because one insists on perturbing around the
trivial, Gaussian point rather than the more physical, non-trivial
fixed point of the renormalization group flow. Interestingly,
thanks to the recent work by Lauscher, Reuter, Percacci, Perini
and others, there is now growing evidence that situation may be
similar with general relativity (see \cite{lr} and references
therein). Impressive calculations have shown that pure Einstein
theory may also admit a non-trivial fixed point. Furthermore, the
requirement that the fixed point should continue to exist in
presence of matter constrains the couplings in physically
interesting ways \cite{pp}.

Let me conclude this introduction with an important caveat.
Suppose one manages to establish that non-perturbative quantum
general relativity (or, supergravity) does exist as a
mathematically consistent theory. Still, there is no a priori
reason to assume that the result would be the `final' theory of
all known physics. In particular, as is the case with classical
general relativity, while requirements of background independence
and general covariance do restrict the form of interactions
between gravity and matter fields and among matter fields
themselves, the theory would not have a built-in principle which
\textit{determines} these interactions. Put differently, such a
theory would not be a satisfactory candidate for unification of
all known forces. However, just as general relativity has had
powerful implications in spite of this limitation in the classical
domain, quantum general relativity should have qualitatively new
predictions, pushing further the existing frontiers of physics. In
section \ref{s3} we will see an illustration of this possibility.

\section{QUANTUM RIEMANNIAN GEOMETRY}
\label{s2}

In a recent short review \cite{aanjp} I have provided a
semi-qualitative description of the quantum Riemannian geometry.
To complement that discussion, in this section I will provide a
concise but more mathematical summary.

The starting point of LQG is a Hamiltonian formulation of general
relativity based on spin connections \cite{aa}. Since all other
basic forces of nature are also described by theories of
connections, this formulation naturally leads to an unification of
all four fundamental forces at a \emph{kinematical} level.
Specifically, the phase space of general relativity is the same as
that of a Yang-Mills theory. The difference lies in dynamics:
whereas in the standard Yang-Mills theory the Minkowski metric
features prominently in the definition of the Hamiltonian, there
are no background fields whatsoever once gravity is switched on.

Let us focus on the gravitational sector of the theory. Then, the
phase space $\ps_\grav$ consists of canonically conjugate pairs
$(A_a^i, \P_{ab}^i)$, where $A_a^i$ is a connection on a
3-manifold $\M$ and $\P_{ab}^i$ a 2-form, both of which take
values in the Lie-algebra $\su(2)$. The connection $A$ enables one
to parallel transport chiral spinors (such as the left handed
fermions of the standard electro-weak model) along curves in $\M$.
Its curvature is directly related to the electric and magnetic
parts of the space-time \emph{Riemann tensor}. The dual $P^a_i$ of
$\P_{ab}^i$ plays a double role.%
\footnote{$P^a_i$ is a vector density, defined via\,\, $3\int_M
\P_{[ab}^i\omega_{c] i} = \int_M P^a_i \omega_a^i$ for any 1-form
$\omega$ on $M$.}
Being the momentum canonically conjugate to $A$, it is analogous
to the Yang-Mills electric field. In addition, $E^a_i := 8\pi G\g
P^a_i$, has the interpretation of an orthonormal triad of frame
field (with density weight 1) on $\M$, where $\g$ is the
`Barbero-Immirzi parameter' representing a quantization ambiguity.
Each triad $E^a_i$ determines a positive definite `spatial'
3-metric $q_{ab}$, and hence the Riemannian geometry of $\M$. This
dual role of $P$ is a reflection of the fact that now $\SU(2)$ is
the (double cover of the) group of rotations of the orthonormal
spatial triads on $\M$ itself rather than of rotations in an
`internal' space associated with $\M$.

To pass to quantum theory, one first constructs an algebra of
`elementary' functions on $\ps_\grav$ (analogous to the phase
space functions $x$ and $p$ in the case of a particle) which are
to have unambiguous operator analogs. The holonomies
\be h_e(A) := {\cal {P}}\, \exp\, \left(-\textstyle{\int}_{e}
{A}\right) \ee
associated with a (piecewise analytic) curve/edge $e$ on $\M$ is a
($\SU(2)$-valued) configuration function on $\ps_\grav$.
Similarly, given a (piecewise analytic) 2-surface $S$ on $\M$, and
a $\su(2)$-valued (test) function $f$ on $\M$,
\be P_{S,f} := \textstyle{\int}_S \Tr\, (f \P) \ee
is a momentum-function on $\ps_\grav$, where $\Tr$ is over the
$\su(2)$ indices.
%
%
The symplectic structure on $\ps_\grav$ enables one to calculate
the Poisson brackets $\{h_e\, , P_{S,f}\}$. The result is a linear
combination of holonomies and can be written as a Lie derivative,
\be \{h_e, \,  P_{S,f}\} = \mathcal{L}_{X_{S,f}}\, h_e \, ,\ee
where $X_{S,f}$ is a derivation on the ring generated by holonomy
functions, and can therefore be regarded as a vector field on the
configuration space $\A$ of connections. This is a familiar
situation in classical mechanics of systems whose configuration
space is a manifold. Functions $h_e$ and vector fields $X_{S,f}$
generate a Lie algebra. As in quantum mechanics on manifolds, the
first step is to promote this algebra to a quantum algebra by
demanding that the commutator be given by $i\hbar$ times the Lie
bracket. The result is a $\star$-algebra $\a$, analogous to the
algebra generated by operators $\exp i\lambda\hat{x}$ and
$\hat{p}$ in quantum mechanics. By exponentiating also the
momentum operators $\hat{P}_{S,f}$ one obtains $\W$, the analog of
the quantum mechanical Weyl algebra generated by $\exp
i\lambda\hat{x}$ and $\exp i\mu\hat{p}$.

The main task is to obtain the appropriate representation of these
algebras. In that representation, \emph{quantum} Riemannian
geometry can be probed through the momentum operators
$\hat{P}_{S,f}$, which stem from classical orthonormal triads. As
in quantum mechanics on manifolds or simple field theories in flat
space, it is convenient to divide the task into two parts. In the
first, one focuses on the algebra $\C$ generated by the
configuration operators $\hat{h}_e$ and finds all its
representations, and in the second one considers the momentum
operators $\hat{P}_{S,f}$ to restrict the freedom.

$\C$ is called the holonomy algebra. It is naturally endowed with
the structure of an Abelian $C^\star$ algebra (with identity),
whence one can apply the powerful machinery made available by the
Gel'fand theory. This theory tells us that $\C$ determines a
unique compact, Hausdorff space $\Ab$ such that the $C^\star$
algebra of all continuous functions on $\A$ is naturally
isomorphic to $\C$. $\Ab$ is called the Gel'fand spectrum of $\C$.
It has been shown to consist of `generalized connections'
$\bar{A}$ defined as follows: $\bar{A}$ assigns to any oriented
edge $e$ in $\M$ an element $\bar{A}(e)$ of $\SU(2)$ (a
`holonomy') such that $\bar{A}(e^{-1}) = [\bar{A}(e)]^{-1}$; and,
if the end point of $e_1$ is the starting point of $e_2$, then
$\bar{A}(e_1\circ e_2) = \bar{A}(e_1)\cdot \bar{A}(e_2)$. Clearly,
every smooth connection $A$ is a generalized connection. In fact,
the space $\A$ of smooth connections has been shown to be dense in
$\Ab$ (with respect to the natural Gel'fand topology thereon). But
$\Ab$ has many more `distributional elements'. The Gel'fand theory
guarantees that every representation of the $C^\star$ algebra $\C$
is a direct sum of representations of the following type: The
underlying Hilbert space is $\H = L^2(\Ab, \dd\mu)$ for some
measure $\mu$ on $\Ab$ and (regarded as functions on $\Ab$)
elements of $\C$ act by multiplication. Since there are many
inequivalent measures on $\Ab$, there is a multitude of
representations of $\C$. A key question is how many of them can be
extended to representations of the full algebra $\a$ (or $\W$)
without having to introduce any `background fields' which would
compromise diffeomorphism covariance. Quite surprisingly, the
requirement that the representation be cyclic with respect to a
state which is invariant under the action of the group of
(piecewise analytic) diffeomorphisms on $\M$ singles out a
\emph{unique} irreducible representation. This result was recently
established for $\a$ by Lewandowski, Oko\l\'ow, Sahlmann and
Thiemann \cite{lost}, and for $\W$ by Fleischhack \cite{cf2}. It
is the quantum geometry analog to the seminal results by Segal and
others that characterized the Fock vacuum in Minkowskian field
theories. However, while that result assumes not only Poincar\'e
invariance but also specific (namely free) dynamics, it is
striking that the present uniqueness theorems make no such
restriction on dynamics. The requirement of diffeomorphism
invariance is surprisingly strong and makes the `background
independent' quantum geometry framework surprisingly tight.

This unique representation was in fact introduced already in the
mid-nineties \cite{al,jb1,mm} and has been extensively used in LQG
since then. The underlying Hilbert space is given by $\H =
L^2(\Ab, \dd\mu_o)$ where $\mu_o$ is a diffeomorphism invariant,
faithful, regular Borel measure on $\Ab$, constructed from the
normalized Haar measure on $\SU(2)$. Typical quantum states can be
visualized as follows. Fix: $(i)$ a graph $\alpha$ on $\M$, and,
$(ii)$ a smooth function $\psi$ on $[\SU(2)]^n$. Then, the
function
\be\label{Psi} \Psi_\g(\bar{A}) := \psi(\bar{A}(e_1),\,\ldots \,
\bar{A}(e_n))\ee
on ${\bar{\cal A}}$ is an element of $\H$. Such states are said to
be \emph{cylindrical} with respect to the graph $\alpha$ and their
space is denoted by $\cyl_\alpha$. These are `typical states' in
the sense that $\cyl := \cup_\alpha\, \cyl_\alpha$ is dense in
$\H$. Finally, as ensured by the Gel'fand theory, the holonomy (or
configuration) operators $\hat{h}_e$ act just by multiplication.
The momentum operators $\hat{P}_{S,f}$ act as Lie-derivatives:
$\hat{P}_{S,f}\, \Psi = -i\hbar\, {\cal L}_{X_{S,f}}\, \Psi$.

Given any graph $\alpha$ in $\M$, and a labelling of each of its
edges by a non-trivial irreducible representation of $\SU(2)$
(i.e., by a non-zero half integer $j$), one can construct a
\emph{finite} dimensional Hilbert space $\H_{\alpha, \vec{j}}$
which can be thought of as the state space of a spin system
`living on' the graph $\alpha$. The full Hilbert space admits a
simple decomposition: $\H = \oplus_{\alpha, \vec{j}}\, \H_{\alpha,
\vec{j}}$. This is called the spin-network decomposition
\cite{rs1,jb2}. The geometric operators discussed in Rovelli's
talk leave each $\H_{\alpha, \vec{j}}$ invariant. Therefore, the
availability of this decomposition greatly simplifies the task of
analyzing their properties \cite{rs2,al5,alrev}.

Key features of this representation which distinguish it from,
say, the standard Fock representation of the quantum Maxwell field
are the following. While the Fock representation of photons makes
a crucial use of the background Minkowski metric, the above
construction is manifestly `background independent'. Second, the
connection itself is not represented as an operator (valued
distribution). Holonomy operators, on the other hand, are
well-defined. Finally, and most importantly, the Hilbert space
$\H$ and the associated holonomy and (smeared) triad operators
only provide a \emph{kinematical} framework ---the quantum analog
of the phase space. Thus, while elements of the Fock space
represent physical states of photons, elements of $\H$ are
\emph{not} the physical states of LQG. Rather, like the classical
phase space, the kinematic setup provides a home for
\emph{formulating} quantum dynamics. In the Hamiltonian framework,
the dynamical content of any background independent theory is
contained in its constraints. In quantum theory, the Hilbert space
$\H$ and the holonomy and (smeared) triad operators thereon
provide the necessary tools to write down quantum constraint
operators. The physical states are solutions to these quantum
constraints. Thus, to complete the program, one has to: i) obtain
the expressions of the quantum constraints; ii) solve the
constraint equations; iii) construct the physical Hilbert space
from the solutions (e.g. by the group averaging procedure); and
iv) extract physics from these physical sectors (e.g., by
analyzing the expectation values, fluctuations of and correlations
between Dirac observables). While strategies have been developed
---particularly through Thiemann's `Master constraint program'
\cite{ttmc}--- to complete these steps, important open issues
remain in the full theory. However, as section \ref{s3}
illustrates, the program has been completed in mini and midi
superspace models, leading to surprising insights and answers to
some long-standing questions.

A more detailed treatment of quantum geometry along the lines
presented here can be found in, e.g., \cite{alrev}.

\section{APPLICATION: HOMOGENEOUS ISOTROPIC COSMOLOGY}
\label{s3}

As emphasized in Sec. \ref{s1}, a central feature of general
relativity is its encoding of the gravitational field in the
Riemannian geometry of space-time. This encoding is directly
responsible for the most dramatic ramifications of the theory: the
big-bang, black holes and gravitational waves. However, it also
leads one to the conclusion that space-time itself must end and
classical physics must come to a halt at the big-bang and black
hole singularities. A central question is whether the situation
improves when gravity is treated quantum mechanically. Analysis of
models within LQG strongly suggests that the answer is in the
affirmative. Because of space limitation, I will restrict myself
to the big bang singularity and that too only in the simplest
setting of homogeneous, isotropic cosmology.

Let us begin with a short list of long-standing questions that any
satisfactory quantum gravity theory is expected to answer:
\begin{quote}
\noindent$\bullet$ How close to the Big Bang does a smooth
space-time of general relativity make sense? In particular, can
one show from first principles that this approximation is
valid at the onset of inflation?\\
$\bullet$ Is the Big-Bang singularity naturally resolved by
quantum gravity? Or, is some external input such as a new
principle or a boundary condition at the Big Bang essential?\\
$\bullet$ Is the quantum evolution across the `singularity'
deterministic? Since one needs a fully non-perturbative framework
to answer this question in the affirmative, in the Pre-Big-Bang
\cite{pbb1} and Ekpyrotic/Cyclic \cite{ekp1,ekp2} scenarios, for
example, so far the answer is in the negative.\\
$\bullet$ If the singularity is resolved, what is on the `other
side'? Is there just a `quantum foam', far removed from any
classical space-time, or, is there another large, classical
universe?
\end{quote}
\noindent For many years, these and related issues had been
generally relegated to the `wish list' of what one would like the
future, satisfactory quantum gravity theory to eventually address.
However, Since LQG is a background independent, non-perturbative
approach, it is well-suited to address them. Indeed, starting with
the seminal work of Bojowald some five years ago \cite{mb1},
notable progress has been made in the context of symmetry reduced,
minisuperspaces.  In this section I will summarize the state of
the art, emphasizing recent developments. For a comprehensive
review of the older work see, e.g., \cite{mbrev}.

Consider the spatially homogeneous, isotropic, $k\!\!=\!\!0$
cosmologies with a massless scalar field. It is instructive to
focus on this model because \emph{every} of its classical
solutions has a singularity. There are two possibilities: In one
the universe starts out at the big bang and expands, and in the
other it contracts into a big crunch. The question is if this
unavoidable classical singularity is naturally tamed by quantum
effects. This issue can be analyzed in the geometrodynamical
framework used in older quantum cosmology. Unfortunately, the
answer turns out to be in the negative. For example, if one begins
with a semi-classical state representing an expanding classical
universe at late times and evolves it back via the Wheeler DeWitt
equation, one finds that it just follows the classical trajectory
into the big bang singularity \cite{aps2,aps3}.

In loop quantum cosmology (LQC), the situation is very different
\cite{aps1,aps2,aps3}. This may seem surprising at first. For, the
system has only a finite number of degrees of freedom and von
Neumann's theorem assures us that, under appropriate assumptions,
the resulting quantum mechanics is unique. The only remaining
freedom is factor-ordering and this is generally insufficient to
lead to qualitatively different predictions. However, for reasons
we will now explain, LQC does turn out to be qualitatively
different from the Wheeler-DeWitt theory \cite{abl}.

Because of spatial homogeneity and isotropy, one can fix a
fiducial (flat) triad $\e^a_i$ and its dual co-triad $\w_a^i$. The
$\SU(2)$ gravitational spin connection $A_a^i$ used in LQG has
only one component $c$ which furthermore depends only on time;
$A_a^i = c\,\, \w_a^i$. Similarly, the triad $E^a_i$ (of density
weight 1) has a single component $p$;\, $E^a_i = p\,(\det \w)\,
\e^a_i$. $p$ is related to the scale factor $a$ via $a^2 = |p|$.
However, $p$ is not restricted to be positive; under $p
\rightarrow -p$ the metric remains unchanged but the spatial triad
flips the orientation. The pair $(c,p)$ is `canonically conjugate'
in the sense that the only non-zero Poisson bracket is given by:
\be \{c,\, p\} = \f{8\pi G \g}{3}\, ,\ee
where as before $\g$ is the Barbero-Immirzi parameter.

Since a precise quantum mechanical framework was not available for
full geometrodynamics, in the Wheeler-DeWitt quantum cosmology one
focused just on the reduced model, without the benefit of guidance
from the full theory. A major difference in LQC is that although
the symmetry reduced theory has only a finite number of degrees of
freedom, quantization is carried out by closely mimicking the
procedure used in \emph{full} LQG, outlined in section \ref{s2}.
Key differences between LQC and the older Wheeler-DeWitt theory
can be traced back to this fact.

Recall that in full LQG diffeomorphism invariance leads one to a
specific kinematical framework in which there are operators
$\h{h}_e$ representing holonomies and $\h{P}_{S,f}$ representing
(smeared) momenta but there is no operator(-valued distribution)
representing the connection $A$ itself \cite{lost,cf2}. In the
cosmological model now under consideration, it is sufficient to
evaluate holonomies along segments $\mu\,\e^a_i$ of straight lines
determined by the fiducial triad $\e^a_i$. These holonomies turn
out almost periodic functions of $c$, i.e. are of the form
$N_{(\mu)}(c) := \exp i\mu (c/2)$. (The $N_{(\mu)}$ are the LQC
analogs of the spin-network functions of LQG.) In the quantum
theory, then, we are led to a representation in which operators
$\h{N}_{(\mu)}$ and $\h{p}$ are well-defined, but there is
\emph{no} operator corresponding to the connection component $c$.
This seems surprising because our experience with quantum
mechanics suggests that one should be able to obtain the operator
analog of $c$ by differentiating $\h{N}_{(\mu)}$ with respect to
the parameter $\mu$. However, in the representation of the basic
quantum algebra that descends to LQC from full LQG, although the
$\h{N}_{(\mu)}$ provide a 1-parameter group of unitary
transformations, the group fails to be weakly continuous in $\mu$.
Therefore one can not differentiate and obtain the operator analog
of $c$.

In quantum mechanics, this would be analogous to having
well-defined (Weyl) operators corresponding to the classical
functions $\exp i\mu x$ but no operator $\h{x}$ corresponding to
$x$ itself. This violates one of the assumptions of the
von-Neumann uniqueness theorem. New representations then become
available which are \emph{inequivalent} to the standard
Schr\"odinger one. In quantum mechanics, these representations are
not of direct physical interest because we need the operator
$\h{x}$. In LQC, on the other hand, full LQG naturally leads us to
a new representation, i.e., to \emph{new quantum mechanics.} This
theory is inequivalent to the Wheeler-DeWitt type theory already
at a kinematical level. In particular, just as we are led to
complete the space $\A$ of smooth connections to the space $\Ab$
of generalized connections in LQG, in LQC we are led to consider
the Bohr compactification $\bar\R_{\rm Bohr}$ of the `$c$-axis'.
The gravitational Hilbert space is now $L^2(\bar{\R}_{\rm Bohr},
\dd\mu_{\rm Bohr})$, rather than the standard $L^2(\R, \dd\mu)$
used in the Wheeler-DeWitt theory \cite{abl} where $\dd\mu_{\rm
Bohr}$ is the LQC analog of the measure $\dd\mu_o$ selected by the
uniqueness results \cite{lost,cf2} in full LQG. While in the
semi-classical regime LQC is well approximated by the
Wheeler-DeWitt theory, important differences manifest themselves
at the Planck scale. These are the hallmarks of quantum geometry
\cite{alrev,mbrev}.

The new representation also leads to a qualitative difference in
the structure of the Hamiltonian constraint operator: the
gravitational part of the constraint is a \emph{difference}
operator, rather than a differential operator as in the
Wheeler-DeWitt theory. The derivation \cite{abl,aps2,aps3} can be
summarized briefly as follows. In the classical theory, the
gravitational part of the constraint is given by $\int d^3x\,
\epsilon^{ijk} e^{-1} E^a_i E^b_j F_{ab\, k}$ where $e = |\det
E|^{1/2}$ and $F_{ab}^k$ the curvature of the connection $A_a^i$.
The part $\epsilon^{ijk} e^{-1} E^a_i E^b_j$ of this operator
involving triads can be quantized \cite{mb1,abl} using a standard
procedure introduced by Thiemann in the full theory \cite{ttbook}.
However, since there is no operator corresponding to the
connection itself, one has to express $F_{ab}^k$ as a limit of the
holonomy around a loop divided by the area enclosed by the loop,
as the area shrinks to zero. Now, quantum geometry tells us that
the area operator has a minimum non-zero eigenvalue, $\Delta$, and
in the quantum theory it is natural to shrink the loop only till
it attains this minimum. There are two ways to implement this idea
in detail (see \cite{abl,aps2,aps3}). In both cases, it is the
existence of the `area gap' $\Delta$ that leads one to a
difference equation. So far, most of the LQC literature has used
the first method \cite{abl,aps2}. In the resulting theory, the
classical big-bang is replaced with a quantum bounce with a number
of desirable features. However, it also has one serious drawback:
at the bounce, matter density can be low even for physically
reasonable choices of quantum states (for details, see
\cite{aps2}); i.e. the theory predicts certain departures from
classical general relativity even in the low curvature regime. The
second  and more recently discovered method \cite{aps3} cures this
problem while retaining the physically appealing features of the
first and, furthermore, has a more direct motivation. Due to space
limitation, I will confine myself only to the second method.

Let us represent states as functions $\Psi(v,\phi)$, where $\phi$
is the scalar field and the dimensionless real number $v$
represents geometry. Specifically, $|v|$ is the eigenvalue of the
operator $\hat{V}$ representing volume (essentially the cube of
the scale factor):
\be \hat{V}\ket{v} =   K\,(\f{8\pi\g}{6})^{\f{3}{2}}\,\, |v|
\,\lp^3 \,\ket{v}\quad {\rm where}\quad K=
\f{3\sqrt{3\sqrt{3}}}{2\sqrt{2}}\,\, \ee
Then, the LQC Hamiltonian constraint assumes the form:
\ba \label{hc3} \p^2_\phi \Psi(v,\phi)  &=& \nonumber  [B(v)]^{-1}
\, \left(C^+(v)\, \Psi(v+4,\phi) + C^o(v) \, \Psi(v,\phi)
+C^-(v)\, \Psi(v-4,\phi)\right)\\
&=:& - \Theta \,\Psi(v,\phi) ~ \ea
where the coefficients $C^\pm(v)$, $C^o(V)$ and $B(v)$ are given
by:
\ba C^+(v) &=& \nonumber \f{3\pi K G}{8} \, |v + 2| \,\,\,
\big| |v + 1| - |v +3|  \big|  \\
C^-(v) &=& \nonumber C^+(v - 4) \quad{\rm and}\quad C^o(v) = -
C^+(v) - C^-(v) \\
B(v) &=& \left(\f{3}{2}\right)^3 \, K\,\, |v| \, \bigg| |v +
1|^{1/3} - |v - 1|^{1/3} \bigg|^3  ~. \ea

Now, in each classical solution, $\phi$ is a globally monotonic
function of time and can therefore be taken as the dynamical
variable representing an \emph{internal} clock. In quantum theory
there is no space-time metric, even on-shell. But since the
quantum constraint (\ref{hc3}) dictates how $\Psi(v,\phi)$
`evolves' as $\phi$ changes, it is convenient to regard the
argument $\phi$ in $\Psi(v,\phi)$ as \emph{emergent time} and $v$
as the physical degree of freedom. A complete set of Dirac
observables is then provided by the constant of motion
$\h{p}_\phi$ and operators $\h{v}|_{\phi_o}$ determining the value
of $v$ at the `instant' $\phi=\phi_o$.

Physical states are the (suitably regular) solutions to Eq
(\ref{hc3}). The map $\h\Pi$ defined by $\h\Pi\, \Psi(v, \phi) =
\Psi(-v, \phi)$ corresponds just to the flip of orientation of the
spatial triad (under which geometry remains unchanged); $\h\Pi$ is
thus a large gauge transformation on the space of solutions to Eq.
(\ref{hc3}). One is therefore led to divide physical states into
sectors, each providing an irreducible, unitary representation of
this gauge symmetry. Physical considerations \cite{aps2,aps3}
imply that we should consider the symmetric sector, with
eigenvalue +1 of $\h{\Pi}$.

To endow this space with the structure of a Hilbert space, one can
proceed along one of two paths. In the first, one defines the
action of the Dirac observables on the space of suitably regular
solutions to the constraints and selects the inner product by
demanding that these operators be self-adjoint \cite{aabook}. A
more systematic procedure is the  `group averaging method'
\cite{dm}. The technical implementation \cite{aps2,aps3} of both
these procedures is greatly simplified by the fact that the
difference operator $\Theta$ on the right side of (\ref{hc3}) is
independent of $\phi$ and can be shown to be self-adjoint and
positive definite on the Hilbert space $L^2(\bar{\R}_{\rm Bohr},
B(v) \dd\mu_{\rm Bohr})$.

The final result can be summarized as follows. Since $\Theta$ is a
difference operator, the physical Hilbert space $\Hp$ has sectors
$\H_\epsilon$ which are superselected; $\Hp = \oplus_\epsilon
\H_\epsilon$ with $\epsilon \in (0,2)$. The overall predictions
are insensitive to the choice of a specific sector (for details,
see \cite{aps2,aps3}). States $\Psi (v,\phi)$ in $\H_\epsilon$ are
symmetric under the orientation inversion $\h{\Pi}$ and have
support on points $v= |\epsilon| + 4n$ where $n$ is an integer.
Wave functions $\Psi(v,\phi)$ in a generic sector solve
(\ref{hc3}) and are of positive frequency with respect to the
`internal time' $\phi$: they satisfy the `positive frequency'
square root
\be -i {\partial_\phi\,\Psi} = \sqrt{\Theta}\, \Psi\, .
\label{2}\ee
of Eq (\ref{hc3}). The physical inner product is given by:
\be \ip{\Psi_1}{\Psi_2}\, = \, \sum_{v\in \{|\epsilon|+4n\}}
B(v)\, \bar\Psi_1(v, \phi_o) \Psi_2(v,\phi_o) \ee
and is `conserved', i.e., is independent of the `instant' $\phi_o$
chosen in its evaluation. On these states, the Dirac observables
act in the expected fashion:
\ba \h{p}_\phi \Psi &=& -i\hbar
{\partial_\phi\,\Psi}\nonumber\\
 \h{v}|_{\phi_o}\,\, \Psi (v,\phi) &=& e^{i
\sqrt{\Theta}(\phi-\phi_o)}\, v\, \Psi(v,\phi_o)\ea

To construct semi-classical states and for numerical simulations,
it is convenient to express physical states as linear combinations
of the eigenstates of $\h{p}_\phi$ and $\Theta$. To carry out this
step, it is convenient to consider the Wheeler-DeWitt theory
first. Let us begin with the observation that, for $v \gg 1$,
there is a precise sense \cite{aps3} in which the difference
operator $\Theta$ approaches the Wheeler DeWitt differential
operator $\ul{\Theta}$, given by
\be \label{wdw2} \ul\Theta \Psi(v,\phi) = {12\pi G}\,\, v\p_v
\big(v\p_v\Psi(v,\phi)\big) \ee
Thus, if one ignores the quantum geometry effects, Eq (\ref{hc3})
reduces to the Wheeler-DeWitt equation
\be \label{wdw1} \partial^2_\phi\Psi = -\ul{\Theta}\,\Psi. \ee
Note that the operator $\ul\Theta$ is positive definite and
self-adjoint on the Hilbert space $L^2_s(\R, \ub{B}(v)\dd v)$
where the subscript $s$ denotes the restriction to the symmetric
eigenspace of $\Pi$ and $\ub{B}(v) := Kv^{-1}$ is the limiting
form of $B(v)$ for large $v$. Its eigenfunctions $\ub{e}_k$ with
eigenvalue $\omega^2 (\ge 0)$ are 2-fold degenerate on this
Hilbert space. Therefore, they can be labelled by a real number
$k$:
\be \ub{e}_k(v) := \f{1}{\sqrt{2\pi}} \, e^{ik\ln |v|}\ee
where $k$ is related to $\omega$ via $\omega= \sqrt{12\pi G}|k|$.
They form an orthonormal basis on  $L^2_s(\R, \ub{B}(v)\dd v)$.  A
`general' positive frequency solution to (\ref{wdw1}) can be
written as
\be \label{wdwsol}\Psi(v, \phi) = \int_{-\infty}^{\infty}\, \dd k
\, \t\Psi(k)\, \ub{e}_k(v) e^{i\omega \phi} \ee
for suitably regular $\t\Psi(k)$.

The complete set of eigenfunctions $e_k(v)$ of the discrete
operator $\Theta$ is also labelled by a real number $k$ and
detailed numerical simulations show that $e_k(v)$ are
well-approximated by $\ub{e}_k (v)$ for $v \gg 1$. The eigenvalues
$\omega^2 (k)$ of $\Theta$ are again given by $\omega= \sqrt{12\pi
G}|k|$. Finally, the $e_k(v)$ satisfy the standard orthonormality
relations $<\!e_k|e_k^\prime\!> = \delta(k, k^\prime)$. A physical
state $\Psi(v,\phi)$ can therefore be expanded as:
\be \label{sol} \Psi(v,\phi) = \int_{-\infty}^{\infty}\! \dd k\,
\t\Psi(k)\,\, e_k^{(s)}(v)\,\, e^{i\omega(k)\phi}\,\ee
where $\t\Psi(k)$ is any suitably regular function of $k$, and
$e^{(s)}_k (v)= (1/\sqrt{2}) (e_k(v)+ e_k(-v))$. Thus, as in the
Wheeler-DeWitt theory, each physical state is characterized by a
free function $\t\Psi(k)$. The difference between the two theories
lies in the functional forms of the eigenfunctions $e_k(v)$ of
$\Theta$ and $\ub{e}_k(v)$ of $\ul\Theta$. While $e_k(v)$ is well
approximated by $\ub{e}_k(v)$ for large $v$, the differences are
very significant for small $v$ and they lead to very different
dynamics.

With the physical Hilbert space and a complete set of Dirac
observables at hand, we can now construct states which are
semi-classical at late times ---e.g., now--- and evolve them
numerically `backward in time'. There are three natural
constructions to implement this idea in detail, reflecting the
freedom in the notion of semi-classical states. In all cases, the
main results are the same \cite{aps2,aps3}. Here I will report on
the results obtained using the strategy that brings out the
contrast with the Wheeler DeWitt theory most sharply.
\begin{figure}
  \begin{center}
    \begin{minipage}{2.0in}
      \begin{center}
        \includegraphics[width=7cm,angle=20]{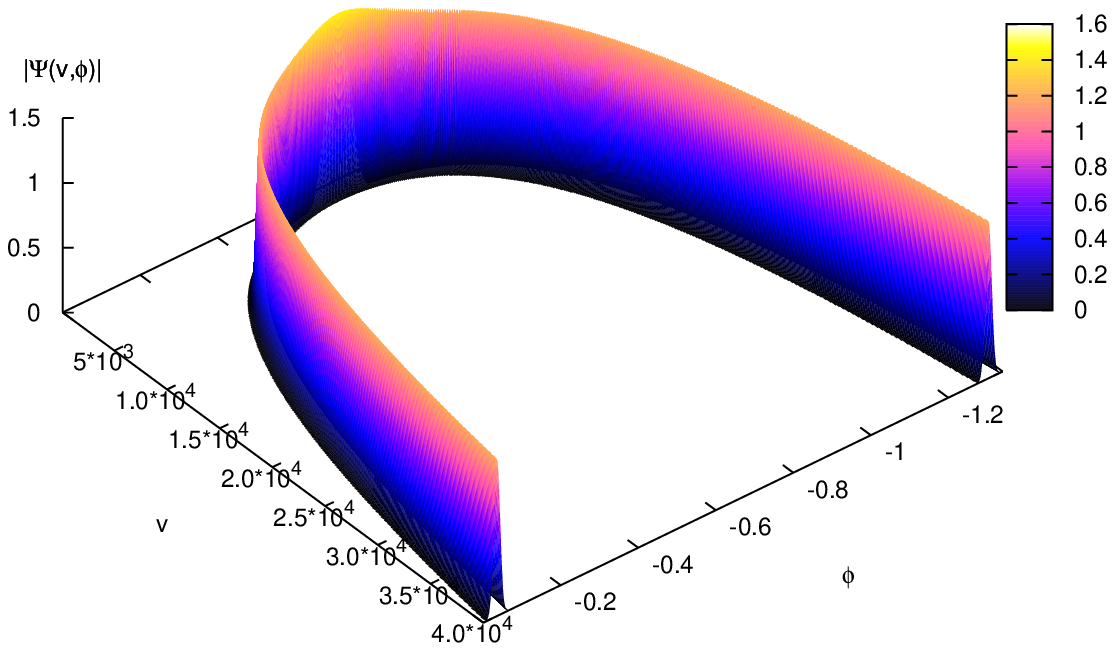}
      \end{center}
    \end{minipage}
    \hspace{2in}
    \begin{minipage}{2.0in}
      \begin{center}\small
        \includegraphics[width=5.5cm,angle=1]{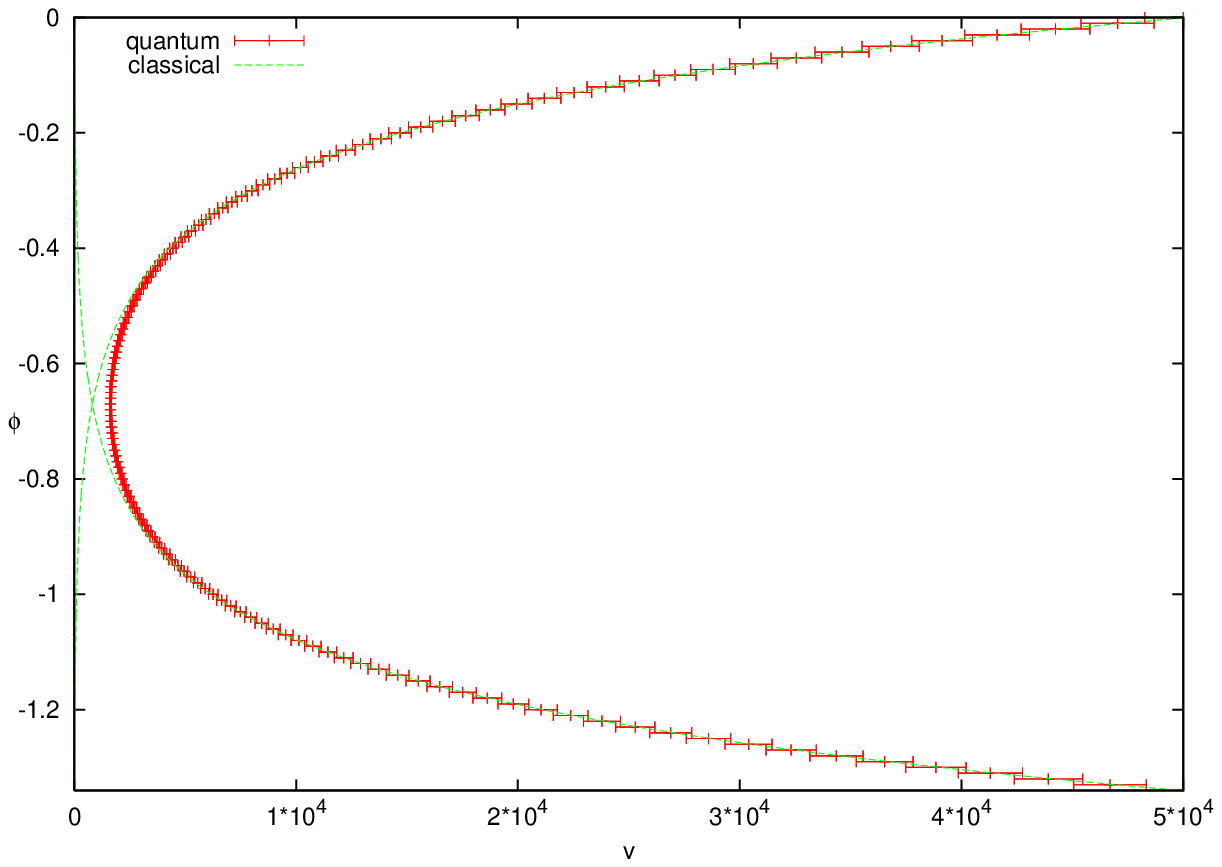}
      \end{center}
    \end{minipage}
    \caption{The figure on left shows the absolute value of the
wave function $\Psi$ as a function of $\phi$ and $v$. Being a
physical state, $\Psi$ is symmetric under $v \rightarrow -v$. The
figure on the right shows the expectation values of Dirac
observables $\h{v}|_{\phi}$ and their dispersions. They exhibit a
quantum bounce which joins the contracting and expanding classical
trajectories marked by fainter lines. In this imulation, the
parameters in the initial data are: $ v^\star = 5\times 10^4,\,\,
p_\phi^\star = 5\times 10^3 \sqrt{G}\hbar$ and  $ \Delta
p_\phi/p_\phi = 0.0025$.}
\end{center}
\end{figure}

As noted before, $p_\phi$ is a constant of motion. For the
semi-classical analysis, we are led to choose a large value
$p_\phi^\star$ ($\gg \sqrt{G}\hbar$). In the closed model, for
example, this condition is necessary to ensure that the universe
can expand out to a macroscopic size. Fix a point $(v^\star,
\phi_o)$ on the corresponding classical trajectory which starts
out at the big bang and then expands, choosing $v^\star \gg 1$. We
want to construct a state which is peaked at $(v^\star,
p_\phi^\star)$ at a `late initial time' $\phi\!=\!\phi_o$ and
follow its `evolution' backward. At `time' $\phi=\phi_o$, consider
then the function
\be \label{sc} \Psi(v, \phi_o) = \int_{-\infty}^\infty \dd k\,
\t\Psi(k)\, \ub{e}_k(v)\, e^{i\omega(\phi_o-\phi^\star)}, \quad
{\rm where}\,\,\, \t\Psi(k) = e^{-\f{(k-k^\star)^2}{2\sigma^2}}
\ee
where $k^\star = - p_\phi^\star/\sqrt{12\pi G\hbar^2}$ and
$\phi^\star = -\sqrt{1/12\pi G}\, \ln (v^\star) +\phi_o$. In the
Wheeler-DeWitt theory on can easily evaluate the integral in the
approximation  $|k^*| \gg 1$ and calculate mean values of the
Dirac observables and their fluctuations. One finds that, as
required, the state is sharply peaked at values $v^\star,
p_\phi^\star$. The above construction is closely related to that
of coherent states in non-relativistic quantum mechanics. The main
difference is that the observables of interest are not $v$ and its
conjugate momentum but rather $v$ and $p_\phi$ ---the momentum
conjugate to `time', i.e., the analog of the Hamiltonian in
non-relativistic quantum mechanics. Now, one can evolve this state
backwards using the Wheeler-DeWitt equation (\ref{wdw1}). It
follows immediately from the form (\ref{wdwsol}) of the general
solution to (\ref{wdw1}) and the fact that $p_\phi$ is large that
this state would remain sharply peaked at the chosen classical
trajectory and simply follow it into the big-bang singularity.

In LQC, we can use the restriction of (\ref{sc}) to points $v =
|\epsilon|+ 4n$ as the initial data and evolve it backwards
numerically. Now the evolution is qualitatively different (see
Fig.1). The state does remains sharply peaked at the classical
trajectory till the matter density reaches a critical value:
\be \rho_{\rm crit} =  \f{\sqrt{3}}{16\pi^2 \gamma^3 G^2 \hbar}\,
, \ee
which is about 0.82 times the Planck density. However, \emph{then
it bounces}. Rather than following the classical trajectory into
the singularity as in the Wheeler-DeWitt theory, the state `turns
around'. What is perhaps most surprising is that it again becomes
semi-classical and follows the `past' portion of a classical
trajectory, again with $p_\phi\! =\! p_\phi^\star$, which was
headed towards the big crunch. Let us we summarize the forward
evolution of the full quantum state. In the distant past, the
state is peaked on a classical, contracting pre-big-bang branch
which closely follows the evolution dictated by Friedmann
equations. But when the matter density reaches the Planck regime,
quantum geometry effects become significant. Interestingly, they
make gravity \emph{repulsive}, not only halting the collapse but
turning it around; the quantum state is again peaked on the
classical solution now representing the post-big-bang, expanding
universe. Since this behavior is so surprising, a very large
number of numerical simulations were performed to ensure that the
results are robust and not an artifact of the special choices of
initial data or of the numerical methods used to obtain the
solution \cite{aps2,aps3}.

For states which are semi-classical at late times, the numerical
evolution in exact LQC can be well-modelled by an effective,
modified Friedman equation :
\be \label{eff} \frac{\dot{a}^2}{a^2}\, =\, \frac{8\pi
G}{3}\,\,\rho\,\, \Big[1 - \frac{\rho}{\rho_{\rm crit}}\Big]
\ee
where, as usual, $a$ is the scale factor. In the limit $\hbar
\rightarrow 0$, $\rho_{\rm crit}$ diverges and we recover the
standard Friedmann equation. Thus the second term is a genuine
quantum correction. Eq. (\ref{eff}) can also be obtained
analytically from (\ref{hc3}) by a systematic procedure \cite{jw}.
But the approximations involved are valid only well outside the
Planck domain. It is therefore surprising that the bounce
predicted by the exact quantum equation (\ref{hc3}) is well
approximated by a naive extrapolation of (\ref{eff}) across the
Planck domain. While there is some understanding of this seemingly
`unreasonable success' of the effective equation (\ref{eff}),
further work is needed to fully understand this issue.

Finally let us return to the questions posed in the beginning of
this section. In the model, LQC has been able to answer all of
them. One can deduce from first principles that classical general
relativity is an excellent approximation till very early times,
including the onset of inflation in standard scenarios. Yet
quantum geometry effects have a profound, global effect on
evolution. In particular, the singularity is naturally resolved
without any external input and there is a classical space-time
also in the pre-big-bang branch. LQC provides a deterministic
evolution which joins the two branches.

\section{DISCUSSION}
\label{s4}

Even though there are several open issues in the formulation of
full quantum dynamics in LQG, detailed calculations in simple
models have provided hints about the general structure. It appears
that the most important non-perturbative effects arise from the
replacement of the local curvature term $F_{ab}^i$ by non-local
holonomies. This non-locality is likely to be a central feature of
the full LQG dynamics. In the cosmological model considered in
section \ref{s3}, it is this replacement of curvature by
holonomies that is responsible for the subtle but crucial
differences between LQC and the Wheeler-DeWitt theory.%
\footnote{Because early presentations emphasized the difference
between $B(v)$ of LQC and $\ub{B}(v) = Kv^{-1}$ of the
Wheeler-DeWitt theory, there is a misconception in some circles
that the difference in quantum dynamics is primarily due to the
non-trivial `inverse volume' operator of LQC. This is not correct.
In deed, in the model considered here, qualitative features of
quantum dynamics, including the bounce, remain unaffected if one
replaces by hand $B(v)$ with $\ub{B}(v)$ in the LQC evolution
equation (\ref{hc3}).}

By now a number of mini-superspaces and a few midi-superspaces
have been studied in varying degrees of detail. In all cases, the
classical, space-like singularities are resolved by quantum
geometry \emph{provided one treats the problem
non-perturbatively.} For example, in anisotropic mini-superspaces,
there is a qualitative difference between perturbative and
non-perturbative treatments: if anisotropies are treated as
perturbations of a background isotropic model, the big-bang
singularity is not resolved while if one treats the whole problem
non-perturbatively, it is \cite{mb-aniso}.

A qualitative picture that emerges is that the non-perturbative
quantum geometry corrections are \emph{`repulsive'}. While they
are negligible under normal conditions, they dominate when
curvature approaches the Planck scale and halt the collapse that
would classically have lead to a singularity. In this respect,
there is a curious similarity with the situation in the stellar
collapse where a new repulsive force comes into play when the core
approaches a critical density, halting further collapse and
leading to stable white dwarfs and neutron stars. This force, with
its origin in the Fermi-Dirac statistics, is \emph{associated with
the quantum nature of matter}. However, if the total mass of the
star is larger than, say, $5$ solar masses, classical gravity
overwhelms this force. The suggestion from LQC is that near Planck
densities a new repulsive force \emph{associated with the quantum
nature of geometry} may come into play which is strong enough to
prevent the formation of singularities irrespective of how large
he mass is. Since this force is negligible until one enters the
Planck regime, predictions of classical relativity on the
formation of trapped surfaces, dynamical and isolated horizons
\cite{akrev} would still hold. But assumptions of the standard
singularity theorems would be violated. There may be no
singularities, no abrupt end to space-time where physics stops.
Non-perturbative, background independent quantum physics could
continue.

The major weakness of the current status of LQG is that so far one
has been able to obtain detailed dynamical predictions only in
symmetry reduced models. These results do provide valuable hints
for the full theory but a large number of ambiguities still remain
there. A fascinating question is whether the singularity
resolution due to quantum geometry is a rather general feature
which is largely insensitive to these ambiguities \cite{nd}. When
matter satisfies the appropriate energy conditions in general
relativity, the Raychaudhuri equation captures the attractive
nature of gravity in a particularly convenient fashion, providing
a central ingredient to the singularity theorems. Is there a
general equation in \emph{quantum geometry} which implies that
gravity effectively becomes repulsive near generic space-like
singularities, thereby halting the classical collapse? If so, one
could construct robust arguments, establishing general
`singularity resolution theorems' for broad classes of situations
in quantum gravity, without having to analyze models, one at a
time.

\vfill\break

\textbf{Acknowledgments:} I would like to thank Martin Bojowald,
Jerzy Lewandowski, and especially Tomasz Pawlowski and Parampreet
Singh for collaboration and numerous discussions. This work was
supported in part by the NSF grants PHY-0354932 and PHY-0456913,
the Alexander von Humboldt Foundation, the Krammers Chair program
of the University of Utrecht and the Eberly research funds of Penn
State.

\end{document}